\documentclass[conference]{IEEEtran}
\IEEEoverridecommandlockouts
\usepackage{cite}
\usepackage{amsmath,amssymb,amsfonts}
\usepackage{algorithmic}
\usepackage{graphicx}
\usepackage{multicol}
\usepackage{textcomp}
\usepackage{xcolor}
\usepackage[a4paper, total={184mm,239mm}]{geometry}
\def\BibTeX{{\rm B\kern-.05em{\sc i\kern-.025em b}\kern-.08em
    T\kern-.1667em\lower.7ex\hbox{E}\kern-.125emX}}

\usepackage[caption=false, font={bf, footnotesize}]{subfig}
\usepackage[acronym]{glossaries}
\usepackage{hyperref}
\usepackage[capitalise, noabbrev]{cleveref}
\usepackage{makecell}
\usepackage{refcount}
\usepackage[htt]{hyphenat}
\usepackage{listings}
\usepackage{soul}
\usepackage{threeparttable}
\usepackage{soul}
\usepackage{orcidlink}
\usepackage{placeins}
\usepackage[pscoord]{eso-pic}
\usepackage{textcomp}
\usepackage{xcolor}

\usepackage[per-mode=symbol,per-symbol =/]{siunitx}
\DeclareSIUnit\GE{GE}
\DeclareSIUnit\kGE{\kilo\GE}
\DeclareSIUnit\MGE{\mega\GE}

\usepackage{tikz}

\usepackage{amssymb}
\usepackage{pifont}
\newcommand{\cmark}{\ding{51}}%
\newcommand{\xmark}{\ding{55}}%

\definecolor{ieee-bright-dblue-100}{rgb}{0.0, 0.3828, 0.6055}
\definecolor{ieee-bright-dblue-80}{rgb}{0.0, 0.4883, 0.6797}
\definecolor{ieee-bright-dblue-60}{rgb}{0.3633, 0.6094, 0.7617}
\definecolor{ieee-bright-dblue-40}{rgb}{0.5898, 0.7383, 0.8398}
\definecolor{ieee-bright-dblue-20}{rgb}{0.8906, 0.8984, 0.9219}
\definecolor{ieee-bright-red-100}{rgb}{0.7266, 0.0469, 0.1836}
\definecolor{ieee-bright-red-80}{rgb}{0.832, 0.3164, 0.3281}
\definecolor{ieee-bright-red-60}{rgb}{0.8906, 0.4922, 0.4805}
\definecolor{ieee-bright-red-40}{rgb}{0.9336, 0.6562, 0.6406}
\definecolor{ieee-bright-red-20}{rgb}{0.9688, 0.8203, 0.8125}
\definecolor{ieee-bright-orange-100}{rgb}{0.9961, 0.6367, 0.0}
\definecolor{ieee-bright-orange-80}{rgb}{0.9844, 0.6953, 0.3125}
\definecolor{ieee-bright-orange-60}{rgb}{0.9883, 0.7695, 0.4844}
\definecolor{ieee-bright-orange-40}{rgb}{0.9922, 0.8359, 0.6562}
\definecolor{ieee-bright-orange-20}{rgb}{0.9961, 0.9219, 0.8164}
\definecolor{ieee-bright-yellow-100}{rgb}{0.9961, 0.8164, 0.0}
\definecolor{ieee-bright-yellow-80}{rgb}{0.9961, 0.8477, 0.2148}
\definecolor{ieee-bright-yellow-60}{rgb}{0.9961, 0.875, 0.4492}
\definecolor{ieee-bright-yellow-40}{rgb}{0.9961, 0.9062, 0.6328}
\definecolor{ieee-bright-yellow-20}{rgb}{0.9961, 0.9531, 0.8125}
\definecolor{ieee-bright-lgreen-100}{rgb}{0.4688, 0.7422, 0.125}
\definecolor{ieee-bright-lgreen-80}{rgb}{0.5742, 0.7852, 0.332}
\definecolor{ieee-bright-lgreen-60}{rgb}{0.6875, 0.8398, 0.5039}
\definecolor{ieee-bright-lgreen-40}{rgb}{0.793, 0.8906, 0.6641}
\definecolor{ieee-bright-lgreen-20}{rgb}{0.8945, 0.9414, 0.8281}
\definecolor{ieee-bright-dgreen-100}{rgb}{0.0, 0.5156, 0.2383}
\definecolor{ieee-bright-dgreen-80}{rgb}{0.1641, 0.6055, 0.3867}
\definecolor{ieee-bright-dgreen-60}{rgb}{0.3906, 0.6953, 0.5234}
\definecolor{ieee-bright-dgreen-40}{rgb}{0.6094, 0.8008, 0.6719}
\definecolor{ieee-bright-dgreen-20}{rgb}{0.8047, 0.8945, 0.8359}
\definecolor{ieee-bright-purple-100}{rgb}{0.5938, 0.1133, 0.5898}
\definecolor{ieee-bright-purple-80}{rgb}{0.6992, 0.3281, 0.668}
\definecolor{ieee-bright-purple-60}{rgb}{0.7812, 0.4961, 0.7461}
\definecolor{ieee-bright-purple-40}{rgb}{0.8555, 0.6602, 0.8281}
\definecolor{ieee-bright-purple-20}{rgb}{0.9219, 0.8281, 0.9023}
\definecolor{ieee-bright-lblue-100}{rgb}{0.0, 0.6094, 0.6484}
\definecolor{ieee-bright-lblue-80}{rgb}{0.0, 0.6797, 0.7188}
\definecolor{ieee-bright-lblue-60}{rgb}{0.2109, 0.75, 0.7812}
\definecolor{ieee-bright-lblue-40}{rgb}{0.5469, 0.8242, 0.8438}
\definecolor{ieee-bright-lblue-20}{rgb}{0.7695, 0.918, 0.9219}
\definecolor{ieee-bright-cyan-100}{rgb}{0.0, 0.707, 0.8828}
\definecolor{ieee-bright-cyan-80}{rgb}{0.0, 0.7227, 0.9453}
\definecolor{ieee-bright-cyan-60}{rgb}{0.2656, 0.7812, 0.957}
\definecolor{ieee-bright-cyan-40}{rgb}{0.5547, 0.8438, 0.9688}
\definecolor{ieee-bright-cyan-20}{rgb}{0.7773, 0.9141, 0.9805}
\definecolor{ieee-bright-white-100}{rgb}{0.9961, 0.9961, 0.9961}
\definecolor{ieee-bright-white-80}{rgb}{0.9961, 0.9961, 0.9961}
\definecolor{ieee-bright-white-60}{rgb}{0.9961, 0.9961, 0.9961}
\definecolor{ieee-bright-white-40}{rgb}{0.9961, 0.9961, 0.9961}
\definecolor{ieee-bright-white-20}{rgb}{0.9961, 0.9961, 0.9961}
\definecolor{ieee-dark-red-100}{rgb}{0.5234, 0.1211, 0.2539}
\definecolor{ieee-dark-red-80}{rgb}{0.6445, 0.2812, 0.3828}
\definecolor{ieee-dark-red-60}{rgb}{0.7422, 0.4727, 0.5234}
\definecolor{ieee-dark-red-40}{rgb}{0.832, 0.6445, 0.6758}
\definecolor{ieee-dark-red-20}{rgb}{0.918, 0.8203, 0.832}
\definecolor{ieee-dark-orange-100}{rgb}{0.9062, 0.4648, 0.1328}
\definecolor{ieee-dark-orange-80}{rgb}{0.9648, 0.5664, 0.3164}
\definecolor{ieee-dark-orange-60}{rgb}{0.9766, 0.6758, 0.4805}
\definecolor{ieee-dark-orange-40}{rgb}{0.9844, 0.7773, 0.6523}
\definecolor{ieee-dark-orange-20}{rgb}{0.9922, 0.8789, 0.8125}
\definecolor{ieee-dark-yellow-100}{rgb}{0.9961, 0.7773, 0.1719}
\definecolor{ieee-dark-yellow-80}{rgb}{0.9961, 0.8086, 0.375}
\definecolor{ieee-dark-yellow-60}{rgb}{0.9961, 0.875, 0.4492}
\definecolor{ieee-dark-yellow-40}{rgb}{0.9961, 0.8984, 0.6875}
\definecolor{ieee-dark-yellow-20}{rgb}{0.9961, 0.9453, 0.8438}
\definecolor{ieee-dark-lgreen-100}{rgb}{0.3945, 0.5508, 0.0938}
\definecolor{ieee-dark-lgreen-80}{rgb}{0.5078, 0.6289, 0.293}
\definecolor{ieee-dark-lgreen-60}{rgb}{0.6367, 0.7188, 0.4688}
\definecolor{ieee-dark-lgreen-40}{rgb}{0.7539, 0.8047, 0.6367}
\definecolor{ieee-dark-lgreen-20}{rgb}{0.875, 0.9023, 0.8125}
\definecolor{ieee-dark-dgreen-100}{rgb}{0.0, 0.3867, 0.2539}
\definecolor{ieee-dark-dgreen-80}{rgb}{0.1836, 0.5, 0.3906}
\definecolor{ieee-dark-dgreen-60}{rgb}{0.3984, 0.6172, 0.5273}
\definecolor{ieee-dark-dgreen-40}{rgb}{0.5938, 0.7422, 0.6758}
\definecolor{ieee-dark-dgreen-20}{rgb}{0.793, 0.8711, 0.8359}
\definecolor{ieee-dark-purple-100}{rgb}{0.4648, 0.1445, 0.5117}
\definecolor{ieee-dark-purple-80}{rgb}{0.5898, 0.3242, 0.6016}
\definecolor{ieee-dark-purple-60}{rgb}{0.6914, 0.4883, 0.6953}
\definecolor{ieee-dark-purple-40}{rgb}{0.7969, 0.6523, 0.793}
\definecolor{ieee-dark-purple-20}{rgb}{0.8945, 0.8203, 0.8945}
\definecolor{ieee-dark-cyan-100}{rgb}{0.0, 0.4492, 0.4648}
\definecolor{ieee-dark-cyan-80}{rgb}{0.0, 0.5469, 0.5664}
\definecolor{ieee-dark-cyan-60}{rgb}{0.3047, 0.6602, 0.668}
\definecolor{ieee-dark-cyan-40}{rgb}{0.5586, 0.7695, 0.7734}
\definecolor{ieee-dark-cyan-20}{rgb}{0.7734, 0.8789, 0.8789}
\definecolor{ieee-dark-dblue-100}{rgb}{0.0, 0.1562, 0.332}
\definecolor{ieee-dark-dblue-80}{rgb}{0.1797, 0.3008, 0.4609}
\definecolor{ieee-dark-dblue-60}{rgb}{0.3828, 0.4609, 0.5859}
\definecolor{ieee-dark-dblue-40}{rgb}{0.5781, 0.6289, 0.7188}
\definecolor{ieee-dark-dblue-20}{rgb}{0.7852, 0.8047, 0.8555}
\definecolor{ieee-dark-grey-100}{rgb}{0.457, 0.4688, 0.4805}
\definecolor{ieee-dark-grey-80}{rgb}{0.5625, 0.5625, 0.5742}
\definecolor{ieee-dark-grey-60}{rgb}{0.6641, 0.6641, 0.6758}
\definecolor{ieee-dark-grey-40}{rgb}{0.7734, 0.7695, 0.7773}
\definecolor{ieee-dark-grey-20}{rgb}{0.8789, 0.8828, 0.8828}
\definecolor{ieee-dark-black-100}{rgb}{0.0, 0.0, 0.0}
\definecolor{ieee-dark-black-80}{rgb}{0.3438, 0.3477, 0.3555}
\definecolor{ieee-dark-black-60}{rgb}{0.5, 0.5078, 0.5195}
\definecolor{ieee-dark-black-40}{rgb}{0.6523, 0.6602, 0.6719}
\definecolor{ieee-dark-black-20}{rgb}{0.8164, 0.8242, 0.8281}

\newcommand*\circnum[1]{\tikz[baseline=(char.base)]{%
            \node[white,shape=circle,fill=ieee-dark-black-100,draw,inner sep=1pt] (char) {\color{ieee-bright-white-100}\sffamily #1};}}

\renewcommand{\baselinestretch}{1.0}

\ifx\showrevision\undefined

\else

    \AddToShipoutPictureFG{%
        \put(%
            8mm,%
            \paperheight-1.5cm%
            ){\vtop{{\null}\makebox[0pt][c]{%
                \rotatebox[origin=c]{90}{%
                    \huge\textcolor{ieee-bright-red-80!75}{\reviewpass}%
                }%
            }}%
        }%
    }
    \AddToShipoutPictureFG{%
        \put(%
            \paperwidth-6mm,%
            \paperheight-1.5cm%
            ){\vtop{{\null}\makebox[0pt][c]{%
                \rotatebox[origin=c]{90}{%
                    \huge\textcolor{ieee-bright-red-80!30}{ETH Zurich - Unpublished - Confidential - Draft - Copyright 2023}%
                }%
            }}%
        }%
    }
\fi

\begin{document}


\title{PELS: A Lightweight and Flexible\\ Peripheral Event Linking System\\ for Ultra-Low Power IoT Processors\\
}


\ifx\blind\undefined
    \author{
        \IEEEauthorblockN{%
        Alessandro Ottaviano\orcidlink{0009-0000-9924-3536}\IEEEauthorrefmark{1}\IEEEauthorrefmark{10},
        Robert Balas\orcidlink{0000-0002-7231-9315}\IEEEauthorrefmark{1}\IEEEauthorrefmark{10}, %
        Philippe Sauter\orcidlink{0000-0000-0000-0000}\IEEEauthorrefmark{1}, %
        Manuel Eggimann\orcidlink{0000-0000-0000-0000}\IEEEauthorrefmark{1}, %
        Luca Benini\orcidlink{0000-0001-8068-3806}\IEEEauthorrefmark{1}\IEEEauthorrefmark{2}%
        }
        \thanks{%
            \IEEEauthorrefmark{10} Both authors contributed equally to this research.
        }
        \IEEEauthorblockA{
            \IEEEauthorrefmark{1}~\textit{Integrated Systems Laboratory, ETH Zurich}, Switzerland \\
            \IEEEauthorrefmark{2}~\textit{Department of Electrical, Electronic, and Information Engineering, University of Bologna}, Italy \\
        }
    }
\else
    \author{%
            \textit{Authors omitted for blind review}
            }
\fi

\maketitle

\begin{abstract}
A key challenge for ultra-low-power (ULP) devices is handling peripheral linking, where the main central processing unit (CPU) periodically mediates the interaction among multiple peripherals following wake-up events. Current solutions address this problem by either integrating event interconnects that route single-wire event lines among peripherals or by general-purpose I/O processors, with a strong trade-off between the latency, efficiency of the former, and the flexibility of the latter. In this paper, we present an open-source, peripheral-agnostic, lightweight, and flexible Peripheral Event Linking System (PELS) that combines dedicated event routing with a tiny I/O processor. With the proposed approach, the power consumption of a linking event is reduced by 2.5 times compared to a baseline relying on the main core for the event-linking process, at a low area of just 7 kGE in its minimal configuration, when integrated into a ULP RISC-V IoT processor.
\end{abstract}


\newacronym{dtm}{DTM}{dynamic thermal management}
\newacronym{stm}{STM}{static thermal management}
\newacronym{hw}{HW}{hardware}
\newacronym{sw}{SW}{software}
\newacronym{ca}{CA}{command/address}
\newacronym{ip}{IP}{intellectual property}
\newacronym{ddr}{DDR}{double data rate}
\newacronym{lpddr}{LPDDR}{low-power double data rate}
\newacronym{rpc}{RPC}{reduced pin count}
\newacronym{dma}{DMA}{direct memory access}
\newacronym{axi}{AXI}{Advanced eXtensible Interface}
\newacronym{dram}{DRAM}{dynamic random access memory}
\newacronym[firstplural=static random access memories (SRAMs)]{sram}{SRAM}{static random access memory}
\newacronym{edram}{eDRAM}{embedded DRAM}
\newacronym[firstplural=systems on chip (SoCs)]{soc}{SoC}{system on chip}
\newacronym{mpsoc}{MPSoC}{multi-processor system on chip}
\newacronym{hesoc}{HeSoC}{heterogeneous system on chip}
\newacronym{sip}{SiP}{system in package}
\newacronym{fpga}{FPGA}{field-programmable gate array}
\newacronym{asic}{ASIC}{application-specific integrated circuit}
\newacronym{phy}{PHY}{physical layer}
\newacronym{ml}{ML}{machine learning}
\newacronym{foss}{FOSS}{free and open source}
\newacronym{cmos}{CMOS}{complementary metal-oxide-semiconductor}
\newacronym{sut}{SUT}{system under test}
\newacronym{isut}{ISUT}{integrated system under test}
\newacronym{rtl}{RTL}{register transfer level}
\newacronym{hil}{HIL}{hardware in the loop}
\newacronym{pil}{PIL}{processor in the loop}
\newacronym{fil}{FIL}{FPGA in the loop}
\newacronym{mil}{MIL}{model in the loop}
\newacronym{sil}{SIL}{software in the loop}
\newacronym{hpc}{HPC}{high performance computing}
\newacronym{mcu}{MCU}{microcontroller unit}
\newacronym{fub}{FUB}{functional unit block}
\newacronym{ecu}{ECU}{electronic control unit}
\newacronym{dcu}{DCU}{domain control unit}
\newacronym{zcu}{ZCU}{zonal control unit}
\newacronym{fame}{FAME}{FPGA Architecture Model Execution}
\newacronym{pl}{PL}{Programmable Logic}
\newacronym{ps}{PS}{Processing System}
\newacronym{apu}{APU}{Application Processing Unit}
\newacronym{ocm}{OCM}{on-chip memory}
\newacronym{pcs}{PCS}{power controller system}
\newacronym{pcf}{PCF}{power control firmware}
\newacronym{plic}{PLIC}{Platform-Level Interrupt Controller}
\newacronym{pmca}{PMCA}{programmable many-core accelerator}
\newacronym{bram}{BRAM}{block RAM}
\newacronym{lut}{LUT}{look-up table}
\newacronym{ff}{FF}{flip-flop}
\newacronym{fsbl}{FSBL}{First Stage BootLoader}
\newacronym{pvt}{PVT}{Process, Voltage, Temperature}
\newacronym{hls}{HLS}{high-level synthesis}
\newacronym{mqtt}{MQTT}{Message Queuing Telemetry Transport}
\newacronym{cots}{COTS}{commercial off-the-shelf}
\newacronym{cpu}{CPU}{central processing unit}
\newacronym{gpu}{GPU}{graphic processing unit}
\newacronym{ibmocc}{IBM OCC}{IBM on-chip controller}
\newacronym{clic}{CLIC}{Core-Local Interrupt Controller}
\newacronym{clint}{CLINT}{Core-Local Interruptor}
\newacronym{scmi}{SCMI}{System Control and Management Interface}
\newacronym{os}{OS}{Operating System}
\newacronym{mimo}{MIMO}{multiple-input multiple-output}
\newacronym{bmc}{BMC}{Baseboard Management Controller}
\newacronym{qos}{QoS}{quality of service}
\newacronym{tdp}{TPD}{thermal design power}
\newacronym{dvfs}{DVFS}{dynamic voltage and frequency scaling}
\newacronym{dfs}{DFS}{dynamic frequency scaling}
\newacronym{dvs}{DVS}{dynamic voltage scaling}
\newacronym{rtu}{RTU}{Real Time Unit}
\newacronym{pe}{PE}{processing element}
\newacronym{noc}{NoC}{network on chips}
\newacronym{pid}{PID}{proportional integral derivative}
\newacronym{sota}{SOTA}{state-of-the-art}
\newacronym{fpu}{FPU}{floating point unit}
\newacronym{pcu}{PCU}{Power Control Unit}
\newacronym{scp}{SCP}{System Control Processor}
\newacronym{mcp}{MCP}{Manageability Control Processor}
\newacronym{occ}{OCC}{On-Chip Controller}
\newacronym{smu}{SMU}{System Management Unit}
\newacronym{ap}{AP}{application-class processor}
\newacronym{vrm}{VRM}{voltage regulator module}
\newacronym{pfct}{PFCT}{periodic frequency control task}
\newacronym{pvct}{PVCT}{periodic voltage control task}
\newacronym{ipc}{IPC}{instructions per cycle}
\newacronym{simd}{SIMD}{single instruction, multiple data}
\newacronym{mctp}{MCTP}{Management Component Transport Protocol}
\newacronym{pldm}{PLDM}{Platform Level Data Model}
\newacronym{rtos}{RTOS}{real-time operating system}
\newacronym{gpos}{GPOS}{general-purpose operating system}
\newacronym{hlc}{HLC}{high-level controller}
\newacronym{llc}{LLC}{low-level controller}
\newacronym{isr}{ISR}{interrupt service routine}
\newacronym{wcet}{WCET}{worst-case execution time}
\newacronym{mcs}{MCS}{mixed criticality system}
\newacronym{isa}{ISA}{instruction set architecture}
\newacronym{csr}{CSR}{Control and Status Register}
\newacronym{apcs}{APCS}{Arm procedure call standard}
\newacronym{fsm}{FSM}{finite state machine}
\newacronym{fiq}{FIQ}{fast interrupt request}
\newacronym{irq}{IRQ}{standard interrupt request}
\newacronym{nvic}{NVIC}{nested vectored interrupt controller}
\newacronym{vic}{VIC}{vectored interrupt controller}
\newacronym{gic}{GIC}{generic interrupt controller}
\newacronym{ge}{GE}{gate equivalent}
\newacronym{icu}{ICU}{interrupt control unit}
\newacronym{srn}{SRN}{service request node}
\newacronym{hart}{HART}{hardware thread}
\newacronym{ir}{IR}{interrupt router}
\newacronym{pc}{PC}{program counter}
\newacronym{intid}{INTID}{interrupt identification number}
\newacronym{csa}{CSA}{context save area}
\newacronym{spm}{SPM}{scratchpad memory}
\newacronym{shv}{SHV}{selective hardware vectoring}
\newacronym{pels}{PELS}{peripheral event linking system}
\newacronym{iot}{IoT}{Internet of Things}
\newacronym{ai}{AI}{artificial intelligence}
\newacronym{vr}{VR}{virtual reality}
\newacronym{ar}{AR}{augmented reality}
\newacronym[firstplural=Advanced Driver Assistance Systems (ADAS)]{adas}{ADAS}{Advanced Driver Assistance System}
\newacronym{v2x}{V2X}{Vehicle-to-Everything}
\newacronym{cps}{CPS}{cyber physical system}
\newacronym{adc}{ADC}{analog-to-digital converter}
\newacronym{clb}{CLB}{configurable logic block}
\newacronym{ulp}{ULP}{ultra-low-power}
\newacronym{ccl}{CCL}{Configurable Custom Logic}
\newacronym{lut}{LUT}{Lookup Table}
\newacronym[firstplural=standard cell memories (SCMs)]{scm}{SCM}{standard cell memory}

\section{Introduction}

The combination of \gls{ulp} and performance in the \gls{iot} market segment leads to demanding \textit{energy-efficiency} requirements for \textit{battery-operated} devices~\cite{EDGE_IOT_SURVEY_2023}, where power savings prolong device autonomy. 
Similarly, in \textit{always-on} wearables such as those found in augmented reality visual processing~\cite{ENVISION}, monitoring services~\cite{SENSOR_FUSION_SURVEY}, and surveillance cameras~\cite{SURVEILLANCE_IOT} energy-efficiency directly translates to reduced cost, and extended device lifetime.
Furthermore, emerging application domains such as in-vehicle communication for automotive demand high computation-per-watt
~\cite{RENESAS_AUTOMOTIVE_GATEWAY} as well as real-time and predictable interaction with a wide range of peripherals and sensors~\cite{SENSOR_FUSION_OBJREC, SENSOR_FUSION_REDUNDANCY} to achieve optimal and fine-grained vehicle control~\cite{SENSOR_FUSION_AUTOMOTIVE}.


To meet the energy requirements of such applications, \glspl{soc} are traditionally designed to reduce standby leakage power when \textit{idle}. This is achieved by enclosing the \textit{processing domain} and the \textit{I/O domain} in different power regions selectively powered on (\textit{active mode}) or off (\textit{sleep mode})~\cite{POWERNAP}.
The \textit{processing domain} only wakes up when a specific condition is detected by the surrounding sensors~\cite{VEGA}.
For such reasons, a typical design goal of \gls{ulp} systems is to minimize the number of wake-up events for the \textit{processing domain}. This not only saves power but also improves the system's predictability by lowering jitter and latency.

A key challenge for ultra-low-power (ULP) devices is handling peripheral linking.
Usually, the processing domain handles peripheral interactions following wake-up events. 
To reduce the number of wake-up events in the processing domain, peripherals can be made more autonomous by enabling inter-peripheral communication (\textit{peripheral event linking}). This communication channel lets peripherals interact directly, potentially bypassing the processing domain.
Peripheral linking can be realized through single-wire event lines (\textit{instant actions}) or with arbitrary commands involving memory-mapped accesses to the peripheral over the system interconnect (hereafter, \textit{sequenced actions}). 
By circumventing the \gls{cpu} and the system interconnect instant actions reduce access latency and minimize jitter for real-time applications.

\begin{figure*} 
    \centering
    \subfloat[\label{fig:linking-irq}]{%
    \includegraphics[width=0.34\linewidth]{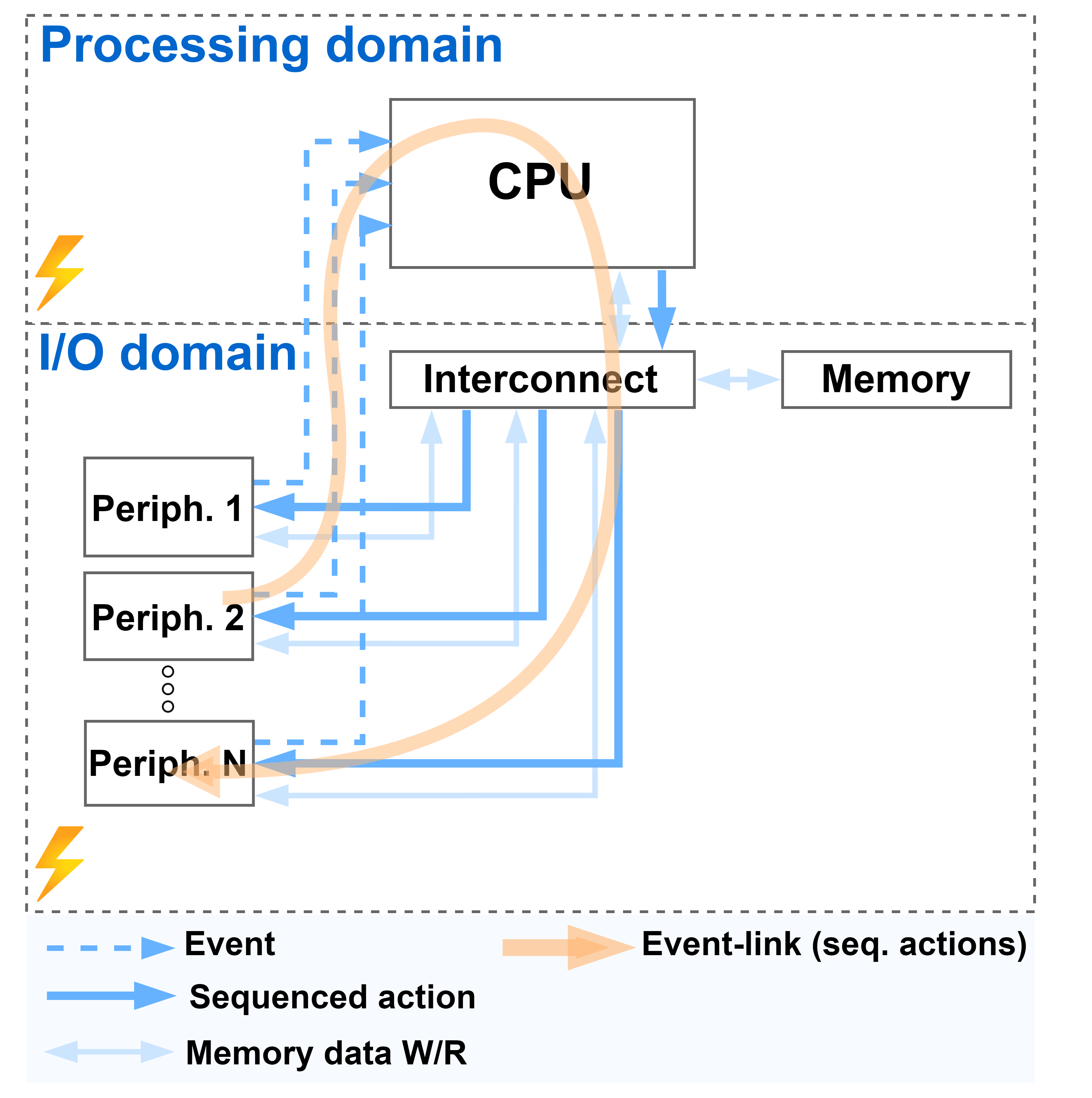}}
    \subfloat[\label{fig:linking-interco}]{%
    \includegraphics[width=0.34\linewidth]{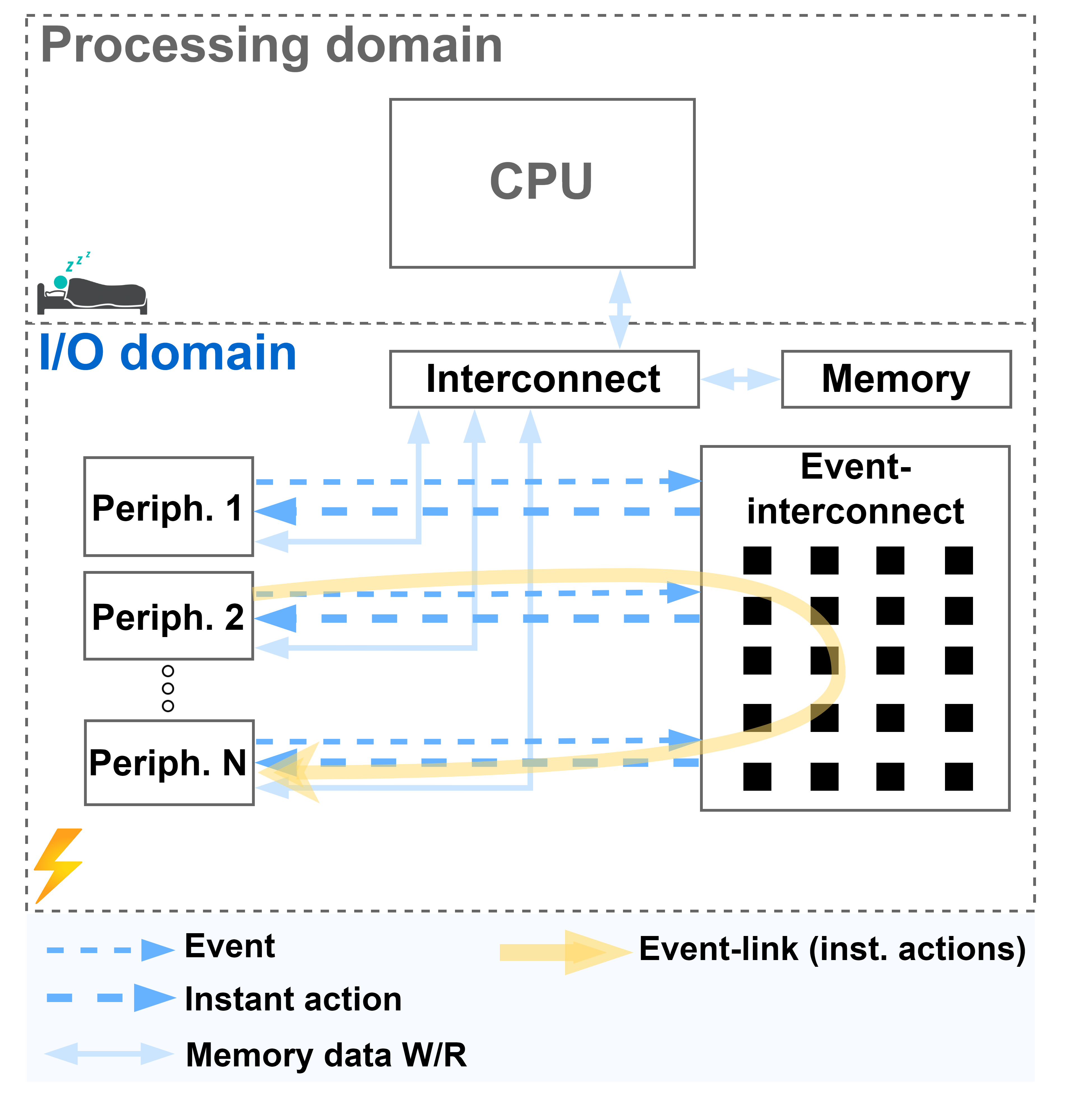}}
    \subfloat[\label{fig:linking-pels}]{%
    \includegraphics[width=0.34\linewidth]{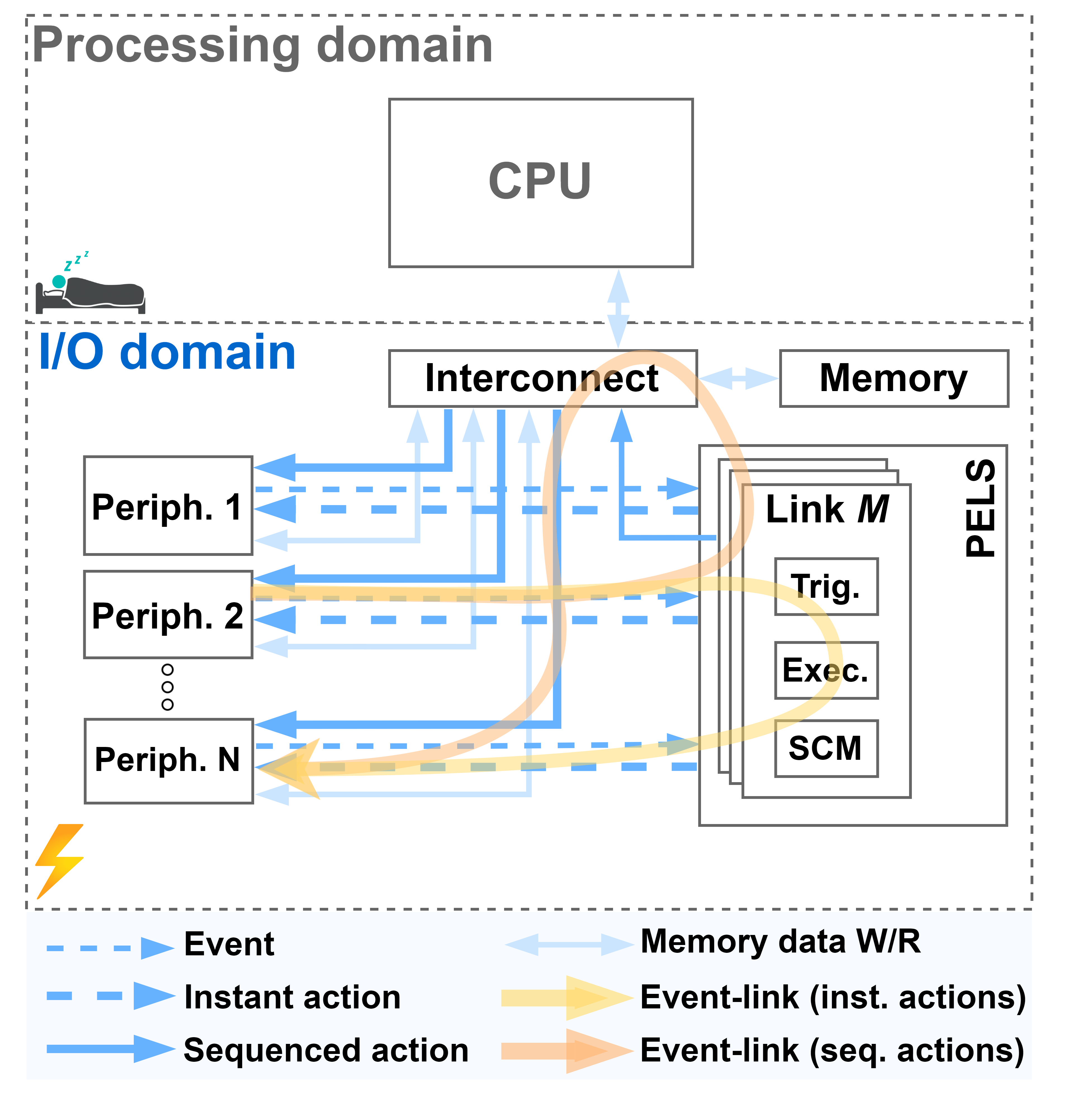}}

    \caption{Profiles of different solutions for handling inter-peripheral communication in \gls{ulp} \glspl{soc}. 
    \textbf{(a)} Traditional software interrupt approach. The main \gls{cpu} is woken up at each inter-peripheral event. 
    \textbf{(b)} Configurable event-interconnect approach with minimal processing capability. The processing domain is bypassed and can sleep, while the event interconnect assures low latency but requires joint codesign with the peripheral system, affecting flexibility. 
    \textbf{(c)} Proposed \gls{pels} design that combines single-wire event lines and sequenced actions through the interconnect. \Gls{pels} doesn't require a priori knowledge of the peripheral system and trades off the low latency and flexibility of existing approaches. 
    (\cref{sec:evaluation}).}
  \label{fig:linking-cases} 
\end{figure*}

For example, consider a periodic timer overflow triggering an \gls{adc} conversion or an SPI end-of-transfer event starting a new data transfer. 
In a traditional system, I/O \gls{dma} engines~\cite{UDMA} can be used for autonomous handling of sensor data movement but are not sufficient to orchestrate inter-peripheral linking, which still requires the main \gls{cpu} to wake up and intervene at each linking event following a classic interrupt mechanism, as depicted in \cref{fig:linking-irq}, adversely affecting the system's response latency, predictability, and power consumption.

Autonomous inter-peripheral event-linking units have been proposed in literature and industry as a complementary solution to I/O \gls{dma} engines to achieve full decoupling between the \textit{processing domain} and the \textit{I/O domain}. 
Existing solutions rely on a peripheral-event interconnect with configurable connections and minimal processing capabilities (\cref{subsec:related_works:periph-event-interco} and \cref{subsec:related_works:periph-event-interco-processing}), as depicted in \cref{fig:linking-interco}. Although area-efficient and low-latency, these approaches require peripheral-aware design to provide a fixed, built-in set of actions. 
Other approaches employ I/O processors in the \textit{I/O domain}. They provide the highest flexibility but incur a higher cost in terms of response latency due to the need to access power-hungry \gls{sram} banks within the \gls{soc} through the system bus (\cref{subsec:related_works:io-processor}).
To the best of the authors' knowledge, a minimally intrusive, low-power, and configurable solution integrating the best of both worlds is missing in the research and industry landscapes.

\textbf{Contribution}
To address these limitations, we propose \gls{pels}, a flexible, lightweight, and open-source event-linking unit. The contributions of this work are:

\begin{itemize}

    \item We design a microcode-based architecture that supports both (i) \textit{instant actions}, i.e., single-wire event lines connecting a \textit{producer} to a \textit{consumer} peripheral, and (ii) \textit{sequenced actions}, i.e., arbitrary commands to control the peripherals through the system interconnect, combining the advantages of existing solutions (\cref{sec:architecture}). Contrary to typical fully-fledged general-purpose processors, \gls{pels} hosts a tiny \gls{scm} to reduce instruction fetch latency and power envelope through the system interconnect bus (\cref{sec:evaluation}).

    \item To the best of the authors' knowledge, \gls{pels} is the first event-linking system available open-source
\ifx\blind\undefined
     ~\footnote{\url{https://www.github.com/pulp-platform/pels}}
\else
     ~\footnote{\texttt{URL omitted for blind review}}
\fi    
     and integrated into a \gls{mcu}-class, open-source, RISC-V \gls{soc}~\cite{PULPISSIMO} targeting \gls{ulp} \gls{iot} applications.

    \item We provide a detailed performance and cost assessment of \gls{pels} in terms of power consumption and area~(\cref{sec:evaluation}). We compare it with a baseline using a traditional interrupt-based mechanism that relies on the main processing core for the event-linking. Furthermore, we show that \gls{pels} remains competitive with \gls{sota}. 
\end{itemize}

\section{Related Works}\label{sec:related_works}

In this section, we describe the main techniques for autonomous peripheral-event handling. We classify the \gls{sota} into three categories. \Cref{tab:related_works:event-linking-approaches} summarizes the main solutions from industry and academia with a feature-based comparison. 

\begin{table*}[t]
    \caption{Feature comparison of existing solutions for autonomous peripheral-event handling. We further highlight whether the analyzed solution is available in the open-source domain.}
    \begin{center}
    \renewcommand{\arraystretch}{1.25} 
    \begin{threeparttable}
    \begin{tabular}{lccccc}
    \rotatebox{0}{\textbf{\thead{System}}} &
    
    \rotatebox{0}{\textbf{\thead{Event Routing \\ topology}}} &
    \rotatebox{0}{\textbf{\thead{Event \\ Processing}}} &
    \rotatebox{0}{\textbf{\thead{Instant \\ Actions}}} &
    \rotatebox{0}{\textbf{\thead{Sequenced \\ Actions}}} &
    \rotatebox{0}{\textbf{\thead{Open \\ source}}} \\

    \hline
    \multicolumn{6}{l}{\textbf{Industry}} \\ 
    \hline
    Silicon Labs PRS~\cite{SILABS_PRS} & Channel & Combinational logic & \cmark & \xmark & \xmark \\ 
    Renesas LELC~\cite{RENESAS_LELC} & Channel & CLB & \cmark & \xmark & \xmark \\ 
    Microchip EVSYS~\cite{MICROCHIP_EVSYS} & Channel & Custom~\tnote{a} & \cmark & \xmark & \xmark \\ 
    Nordic PPI~\cite{NORDIC_PPI} & Channel & Custom~\tnote{b} & \cmark & \xmark & \xmark \\ 
    STMicroelectronics PIM~\cite{STM_PIM} & Matrix & \xmark & \cmark & \xmark & \xmark \\ 
    NXP XGATE~\cite{NXP_XGATE} & --- & Microcode~\tnote{c} & \xmark & \cmark & \xmark \\ 

    \hline
    \multicolumn{6}{l}{\textbf{Academia}} \\ 
    \hline

    AESRN (Bj{\o}rnerud et al.~\cite{BJORN_EVENT}) & Channel & CLB~\tnote{d} & \cmark & \xmark & \xmark \\
    \textbf{This work} & Channel & Microcode & \cmark & \cmark & \cmark \\
    \hline
    \end{tabular}
    \begin{tablenotes}
    \item[a] Up to three events can be routed to the \gls{ccl}, which allows the generation of a new output event as a function of the input events. The function is given by a \gls{lut}.
    \item[b] One channel can trigger up to two actions simultaneously, which can be seen as limited broadcasting.
    \item[c] Designed as a co-processor to take the interrupt load off the main processor.
    \item[d] Uses fully asynchronous logic for processing and routing
    \end{tablenotes}
    \end{threeparttable}
    \label{tab:related_works:event-linking-approaches}
    \end{center}
\end{table*}

\subsection{Software Interrupts}
In a purely \gls{sw}-driven approach, peripheral events are handled and routed through the processing domain using common interrupt handlers (\cref{fig:linking-irq}).
This solution creates dependencies between the \textit{I/O domain} and \textit{processing domain}, which can increase the response latency and power consumption due to the CPU waking up to handle the event processing.

\subsection{Peripheral-event interconnect}\label{subsec:related_works:periph-event-interco}
A common solution to handle inter-peripheral linking in \gls{hw} uses peripheral-event interconnects. In their simplest form, these are a set of \emph{fixed} connections between peripherals to forward events to each other. 
A significantly more flexible solution is to make the peripheral-event interconnect routing configurable. Several routing topologies exist, such as matrix configurations~\cite{STM_PIM} or multiple multiplexer-demultiplexer channels~\cite{SILABS_PRS}\cite{RENESAS_LELC}\cite{MICROCHIP_EVSYS}\cite{NORDIC_PPI}. \Cref{fig:linking-interco} shows a representation of such systems.
While these approaches offer good area and power efficiency, programmability, and predictable as well as low event latency, they lack flexibility as inter-peripheral connections are decided at design time, i.e., the peripheral design has to be tailored to the event interconnect.

\subsection{Processing on interconnect}\label{subsec:related_works:periph-event-interco-processing}
\subsubsection{FPGA approach}
If an application demands complex criteria to trigger an action, the event-interconnect topology might be insufficient, as it still requires the processing system to intervene. To address this limitation, the event linking system can be enhanced with minimal processing capabilities, as in the case of the Peripheral Reflex System~\cite{SILABS_PRS}, or with \glspl{clb} of arbitrary complexity de facto integrating a small \gls{fpga} within the peripheral system.
A common approach is adding a \gls{clb} to the channel topology with minimal inputs and outputs that can also be connected to another channel~\cite{RENESAS_LELC}~\cite{BJORN_EVENT}.

\subsubsection{I/O processor}\label{subsec:related_works:io-processor}
The last approach in this category consists of adding a tiny general-purpose processor into the peripheral system, which can directly access other peripheral devices through the system interconnect in the \textit{I/O domain} as the main core in the \textit{processing domain} would~\cite{NXP_XGATE}.
The latency of solutions based on a general-purpose I/O processor can be limited by (i) the \textit{system interconnect}, as a slow bus affects the latency while the routing arbitration influences the system's predictability; (ii) the \textit{memory}, as using the system's local memory trades off area reuse for latency; (iii) \textit{datapath width}, which should be at least equal to the system interconnect width to complete accesses within one clock cycle.

Compared to \gls{sota}, \gls{pels} allows full programmability of the linking topology with two operation modes.
On the one hand, it allows to trigger instant actions when strict latency constraints are required. 
On the other hand, its microcode design enables flexible integration of new sequenced actions for peripherals as long as they are realizable through the system interconnect. Furthermore, a priori knowledge of the peripheral system design is not required.

\section{Architecture}\label{sec:architecture}
This section describes \gls{pels}'s microarchitecture, depicted in \cref{fig:pels-link-arch}. 
\Gls{pels} collects events from peripherals and triggers the execution of a sequence of \textit{commands} (\cref{subsubsec:archi-commands}), referred to as a \textit{sequenced action}. Hence, it acts as a controller reading and writing to the peripheral devices over the system interconnect bus.
At the same time, it also supports commands triggering peripheral-specific actions through single-wire event lines as simpler peripheral-event handling systems do (\cref{sec:related_works}). These are referred to as \textit{instant actions}. An overview of \gls{pels}'s operation modes and key differences compared to \gls{sota} is provided in \cref{fig:linking-cases}.

\subsubsection{\textbf{Link}}\label{subsubsec:archi-link}
\begin{figure*}
    \centering
    \includegraphics[width=0.75\textwidth]{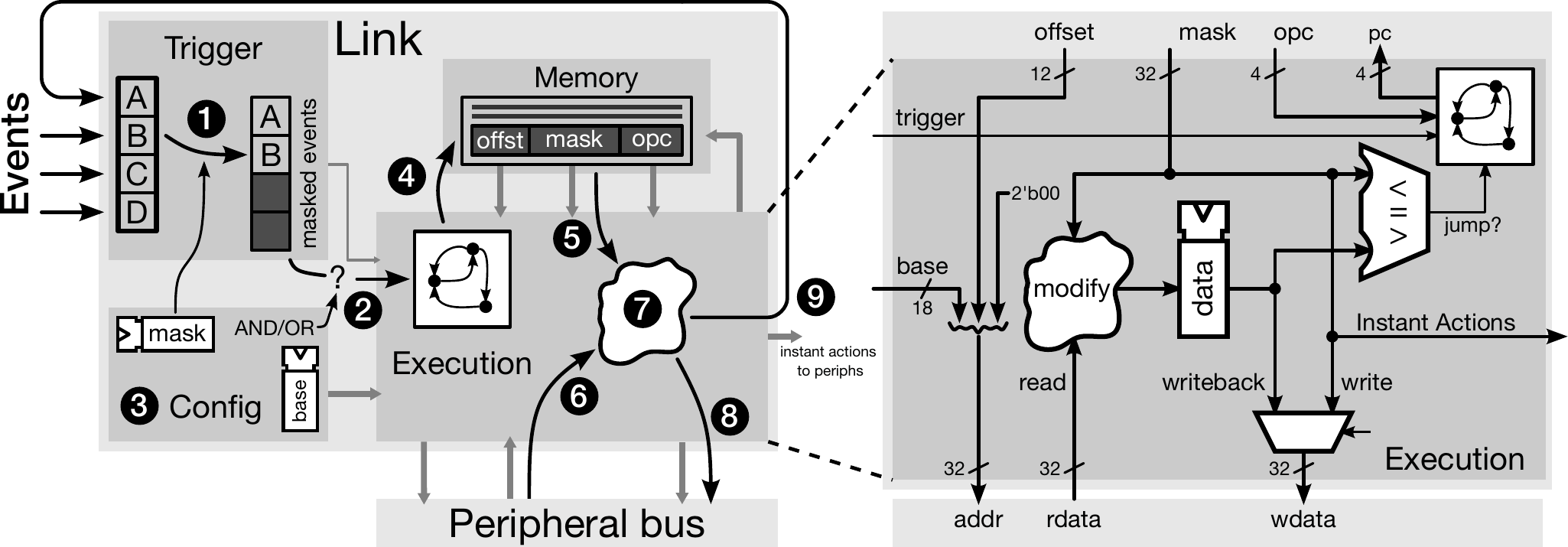}
    \caption{Architectural overview of a single link. The figure highlights the execution unit microarchitecture.}
    \label{fig:pels-link-arch}
\end{figure*}

To provide parallelism when servicing multiple peripheral linking events, \gls{pels} is internally organized into independent linking units, referred to as \emph{links}, as shown in \cref{fig:linking-pels}. When each link executes a sequenced action, the interconnect topology determines the number of parallel interactions with the peripheral devices, while its arbitration policy affects each link's typical and maximum latency, especially in the worst-case scenario where all links try to access peripherals simultaneously.
As shown in \cref{fig:pels-link-arch}, a link consists of three main building blocks: the \textit{trigger unit}, the \textit{execution unit}, and the \textit{\gls{scm}}, described in the following.

\paragraph{\textbf{Trigger unit}}
Input events to \gls{pels} are broadcasted to all links. Each link's trigger unit masks incoming events \circnum{1}, further checked against trigger conditions \circnum{2}, such as \textit{all-selected-active} (\texttt{AND}) or \textit{any-selected-active} (\texttt{OR}). For instance, a trigger condition can be a threshold to generate an event.
The main \gls{cpu} configures both masking and triggering conditions through each link's private configuration registers \circnum{3}.

\paragraph{\textbf{Private \gls{scm}}}
If the condition to propagate the event is satisfied, a trigger signal is generated and buffered with a FIFO to prevent interference with a running execution unit. 
A \gls{fsm} reads line-by-line the microcode to be executed from the integrated instruction memory \circnum{4}. Each line in the memory holds one \textit{command}. Dedicated instruction memory allows for predictable latency due to not having to contend on a shared bus, which can be as low as a single clock cycle when linking with one interacting peripheral.
The instruction memory is implemented with \glspl{scm}, which incurs minimal power consumption and area overhead for small memory footprints, as in the case of \gls{pels}. With a \glspl{sram}, the sense amplifiers would otherwise dominate the area~\cite{SCM}.

\paragraph{\textbf{Execution unit}}
\Cref{fig:pels-link-arch} highlights the microarchitecture of the execution unit.
One clock cycle after a successful triggering condition, the execution unit receives the first command from the instruction memory \circnum{5}. 
In the simplest case, this command triggers an instant action through single-bit event lines directly connected to the peripheral devices.

Otherwise, \gls{pels} can issue sequenced actions for peripherals without dedicated event lines. 
When the command becomes available, the execution unit issues a read operation to the system interconnect, and it stalls until it receives the register's value from the peripheral \circnum{6}. The value is then modified \circnum{7} according to the command. 
One cycle after the read succeeds, the modified value is written back \circnum{8} to the peripheral register. 
Finally, \cref{fig:pels-link-arch} shows that some event lines triggered by instant actions can be looped back to the incoming events, which enables inter-link triggering \circnum{9}.

\subsubsection{\textbf{Commands}}\label{subsubsec:archi-commands}

The primitive \texttt{write} command writes a known value to a register. More complex and higher latency commands read the value from a selected register, perform a bitwise operation, and write the result back to the register (\textit{read-modify-write}). They comprise \texttt{set}, \texttt{clear}, and \texttt{toggle} commands. 
Some use cases for such commands include setting/resetting peripheral register flags or toggling GPIOs.

Read-modify-write commands constrain the microcode encoding, as for every peripheral register access, an address and a mask are needed. 
Provided that a memory transaction for bus-based peripheral interaction requires both the datum and an address, a single-cycle transaction needs more than 32-bit encoding. To reduce the address encoding space, \gls{pels} only requires a word-addressed offset relative to a base address specific to each link. 
Overall, the number of offset bits lies within a range of 10-14 bits, leading to a command encoding with 4-bit \textit{op-code}, a 12-bit field address, and a 32-bit data.

To address the use-case of sensor readout followed by a threshold comparison (a common operation for event-linking, see \cref{fig:pels-pseudocode}), we include a \texttt{capture} and \texttt{jump-if} command. \texttt{capture} performs a masked read and stores it in a single 32-bit register per link. \texttt{jump-if} reads the value from the datapath register and compares it against its 32-bit operand. If the condition is satisfied, it jumps to another command. 

Two additional commands, namely \texttt{loop} and \texttt{wait}, are implemented to enable the design of non-nestable hardware loops and wait counters. In particular, they subsume watchdog-like functions without requiring an external timer.

Finally, the \texttt{action} command is the core of the \textit{instant action} mechanism. It sets or toggles one or multiple outgoing single-wire event lines. The 12-bit field reserved for addresses with sequenced actions is instead used to select a group of event lines, while the 32-bit operand is used to set them. The single-wire event lines enable straightforward integration of any peripherals supporting built-in actions as \gls{sota} approaches shown in \cref{fig:linking-interco}. Additionally, \gls{pels} allows links to trigger each other through specific instant actions. This feature is beneficial when links have limited instruction memory or can access specific peripherals only, and enables links specialization and diversification in handling a subset of peripherals, a feature missing in existing \gls{sota}.

\subsubsection{\textbf{Programming model}}\label{subsubsec:archi-example}

\begin{figure}[t]
	\centering
    \includegraphics[width=.65\columnwidth]{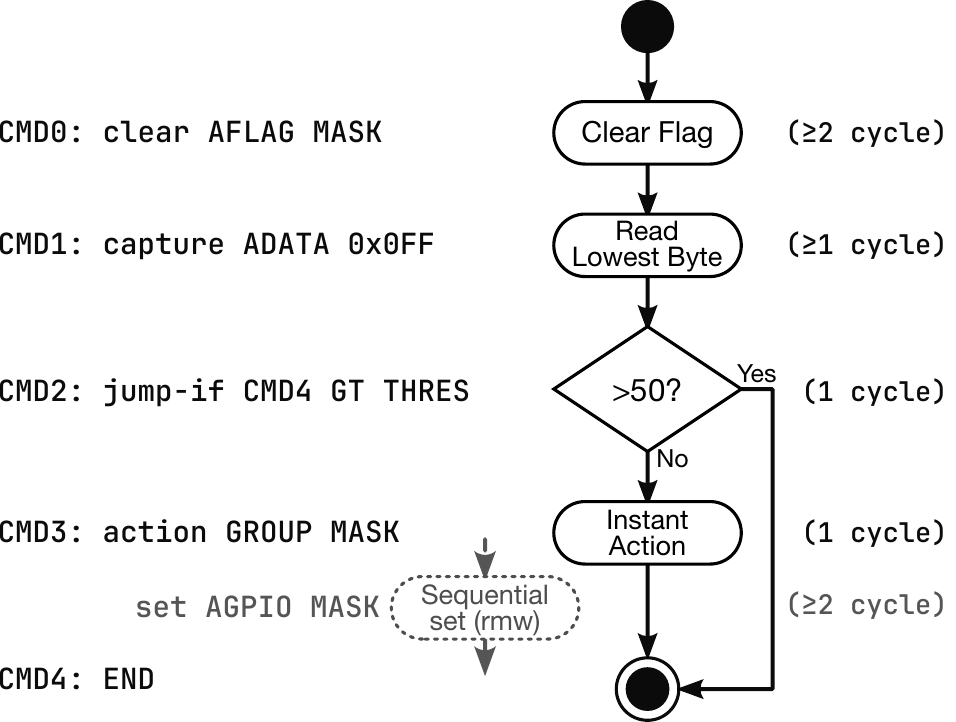}
	\caption{Pseudocode showing \gls{pels}'s flexibility in providing both sequenced and instant actions and associated latency in clock cycles.}
	\label{fig:pels-pseudocode}
\end{figure}


\Cref{fig:pels-pseudocode} provides an example of \gls{pels}'s dual-mode operation flexibility. In the presented pseudocode, \gls{pels} implements a threshold-triggered operation following sensor readout (e.g., a thermistor/varistor) with both instant and sequenced actions.
In the former mode, the output event line is redirected to the peripheral (e.g., a GPIO) if the latter is co-designed to support single-wire event lines.
With sequenced actions instead, \gls{pels} directly controls the peripheral through the peripheral interconnect, regardless of its event interface design. The figure shows the latency of each stage in clock cycles; as discussed in \cref{subsubsec:archi-link}, instant actions have lower, fixed latency but require specialized logic.

\section{Evaluation}\label{sec:evaluation}



%







In this section, we first introduce the RISC-V \gls{mcu} where \gls{pels} is integrated (\cref{subsec:evaluation-pulpissimo}). Then, we evaluate \gls{pels}'s beneficial impact on the \gls{mcu} power consumption when mediating peripheral linking operations (\cref{subsec:evaluation-functional}) and low silicon area overhead (\cref{subsec:evaluation-implementation}), as a result of the design tradeoffs described in the previous section.

\subsection{PULPissimo RISC-V \gls{mcu}}\label{subsec:evaluation-pulpissimo}
PULPissimo~\cite{PULPISSIMO} is a single-core, \gls{mcu}-class RISC-V system designed for embedded \gls{ulp} applications. Among the features it supports to achieve \gls{ulp} while retaining high energy efficiency, we mention: (i) an autonomous I/O \gls{dma} to decouple data collection and processing, (ii) near-threshold operation~\cite{PULP_NEAR_THRESHOLD}, and (iii) adaptive body-biasing to reduce standby leakage power~\cite{POWERNAP}.
We opt for the single issue, in order, 2-pipeline stage Ibex core~\cite{IBEX} among the processors supported by the \gls{mcu} because of its minimal area (\cref{subsec:evaluation-implementation}). 

\begin{figure}
    \centering
    \includegraphics[width=0.8\columnwidth]{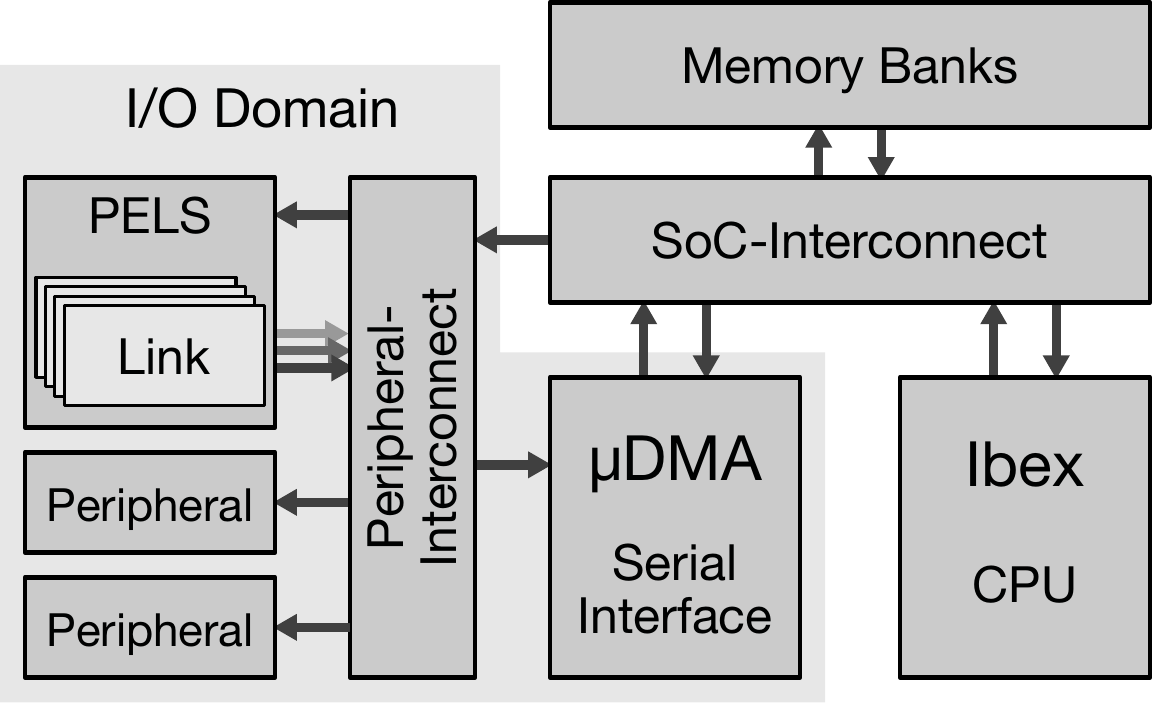}
    \caption{\Gls{pels} integration in PULPissimo \gls{soc}.}
    \label{fig:archi-pels-integration}
\end{figure}

In this work, we extend PULPissimo with \gls{pels} as depicted in \cref{fig:archi-pels-integration}.
The main configuration parameters to tune \gls{pels} performance within a system are the number of links (number of parallel actions) and the memory size (number of commands per link). 
Furthermore, when issuing \textit{sequenced actions}, the topology of the system interconnect and its arbitration policy affect (i) the number of links that can access a group of peripherals in parallel and (ii) the system's predictability due to variable latency. In the present work, we rely on PULPissimo's implementation of round-robin arbiters in the system interconnect to guarantee fair bandwidth distribution. 

\subsection{Functional performance}\label{subsec:evaluation-functional}



To evaluate the architecture's performance, we design an event-linking application consisting of a threshold-crossing check after I/O \gls{dma}-managed sensor readout through the SPI interface, similarly to the example shown in \cref{fig:pels-pseudocode}. %
We compare \gls{pels}'s mediation through sequenced actions with an interrupt-based mechanism redirecting the linking event to the Ibex core in two scenarios: (i) iso-latency, i.e., \gls{pels} and Ibex are independently clocked at different frequencies to meet specific latency requirements and (ii) iso-frequency without latency constraints.
Both evaluations on such benchmarks are carried out with cycle-accurate RTL simulations, and we further estimate the power consumption of a linking event using Synopsys PrimeTime on the synthesized netlist.
We present power estimations for sequenced actions, as instant actions introduce negligible dynamic power.

\begin{figure}
    \centering
    \includegraphics[width=\columnwidth]{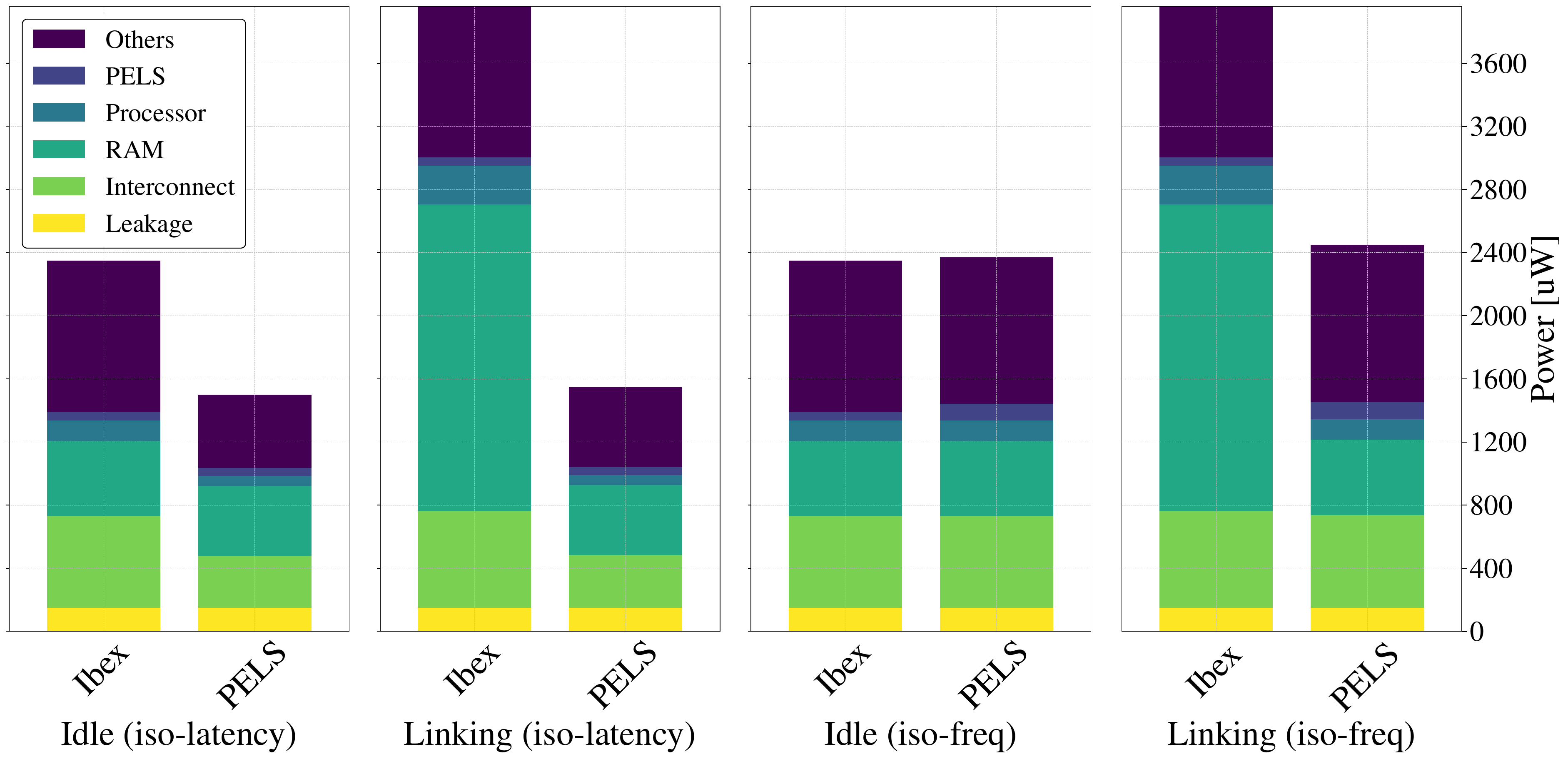}
    \caption{Power estimation with iso-latency and iso-frequency conditions between \gls{pels} and Ibex.}
    \label{fig:eval-pels-power}
\end{figure}

\Cref{fig:eval-pels-power} provides a breakdown of PULPissimo power consumption when waiting for and handling the event linking for both iso-latency and iso-frequency scenarios.
In the iso-latency scenario, \gls{pels} and Ibex match a \SI{500}{\nano \second} latency requirement at \SI{27}{\mega \hertz} and \SI{55}{\mega \hertz} respectively. 
The power consumed during the event linking process in the \gls{soc} is reduced by 2.5$\times$ when handled by \gls{pels} compared to the Ibex interrupt-driven mechanism. 
We argue this is due to the decreased frequency (affecting the dynamic power) to meet the imposed latency constraint and the reduced switching activity around the memory system (3.7$\times$ less power).
Analogously, waiting for an event (\emph{idle} mode) without standby leakage power-saving techniques uses 1.5$\times$ less power with \gls{pels}-driven event linking.

In the iso-frequency case, both systems are clocked at \SI{55}{\mega \hertz}. This scenario also shows the benefits of the event-linking system in the active mode, with a power consumption reduction of 1.6$\times$. As in the previous case, one of the main contributions to the reduced power envelope comes from the little activity around the memory system in the \gls{pels}-driven scenario (4.3$\times$ less power draw). %

For a latency comparison, we observe 7 and 2 clock cycles for sequenced and instant actions, respectively. The former is slower and depends on the peripheral bus protocol (APB for~\cite{PULPISSIMO}), while the latter is fixed. 
As anticipated, Ibex takes significantly longer (16 clock cycles) to handle the same linking event because of the interrupt handling mechanism overhead.

\subsection{Implementation}\label{subsec:evaluation-implementation}
\begin{figure}[tb] 
    \centering
    \subfloat[\label{fig:eval-pels-area-sweep}]{%
    \includegraphics[width=\columnwidth]{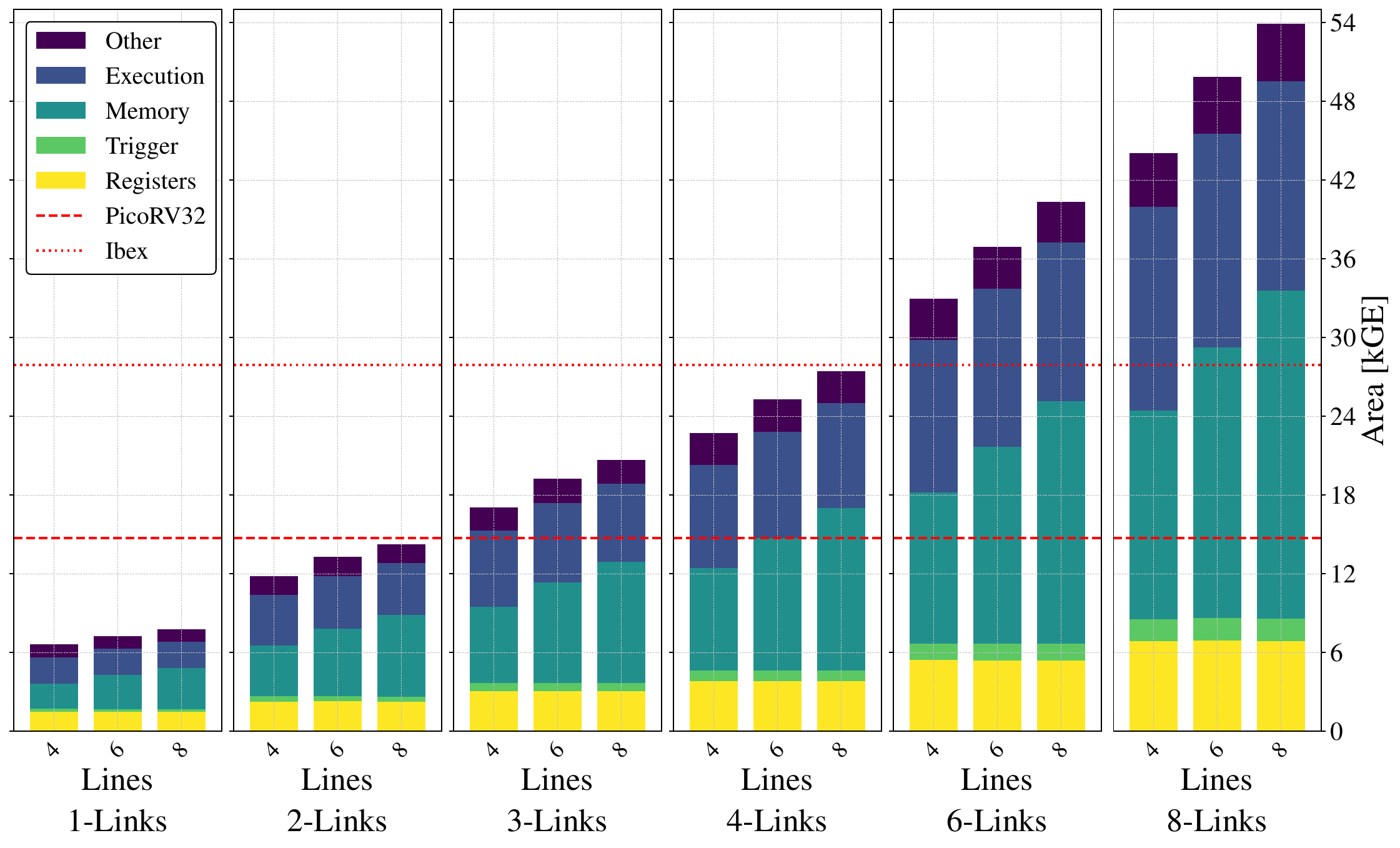}}

    \subfloat[\label{fig:eval-pels-area-pulpissimo}]{%
    \includegraphics[width=\columnwidth]{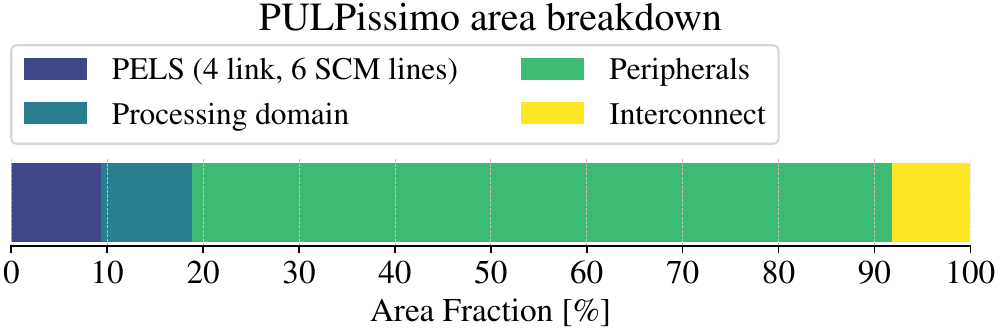}}

    \caption{\Gls{pels} silicon area evaluation in TSMC's \SI{65}{\nano\meter} technology, \SI{250}{\mega\hertz}, TT corner, \SI{25}{\celsius}. \textbf{(a)} \Gls{pels} area sweep over the number of links and \gls{scm} lines. \textbf{(b)} \Gls{pels} area overhead in PULPissimo.}
  \label{fig:eval-pels-area} 
\end{figure}

We synthesize \gls{pels} in several configurations and compare it against extremely low-footprint processors that could serve as fully-fledged general-purpose solutions for event-linking, proving the benefits of the proposed microcode approach. 
Topological synthesis is carried on with Synopsys Design Compiler 2022.03, targeting TSMC's \SI{65}{\nano\meter} technology at \SI{250}{\mega\hertz}, TT corner, and \SI{25}{\celsius}.
We vary the number of links from 1 to 8, sufficient to cover realistic use cases. For each link, we consider \glspl{scm} with 4, 6, and 8 lines (microcode commands).
\Cref{fig:eval-pels-area-sweep} shows the results. 
\Gls{pels} compares favorably with some of the leanest general-purpose cores in the RISC-V landscape, represented as dashed horizontal lines in the figure, that we synthesize with equal frequency and \gls{pvt} conditions for a fair comparison. 
In its minimal configuration (1 link, 4 commands), \gls{pels} results in an area of about \SI{7}{\kGE}, 4$\times$ smaller than Ibex~\footnote{\url{https://github.com/lowRISC/ibex}} (about \SI{27}{\kGE}) and 2$\times$ smaller than PicoRV32~\footnote{\url{https://github.com/YosysHQ/picorv32}} (about \SI{14.5}{\kGE}), without considering the external SRAMs that those RISC architectures require for load and store operations.

\Gls{pels} proves to be lightweight also within the RISC-V \gls{soc} it is integrated into.
\Cref{fig:eval-pels-area-pulpissimo} shows that a \textit{4-links} \gls{pels} accounts for about 9.5\% PULPissimo's area (1\% when considering the SRAM - \SI{192}{\kibi \byte} in the implemented configuration).   

The presented evaluation shows that \gls{pels} is highly competitive in terms of silicon area, power consumption, and latency compared to both general-purpose solutions and existing \gls{sota} addressing event-linking with bespoke event interconnects. 
Instead, \gls{pels} resolves these limitations by combining \emph{instant} and \emph{sequenced} actions with an \gls{scm}-based memory system, guaranteeing flexibility and fast access to private memory. 

\section{Conclusion}
In this paper, we address the challenge of designing flexible near-sensor processing systems that can perform fully autonomous multi-sensor and inter-peripheral communication.
To achieve this goal, we propose a peripheral-agnostic, fully autonomous, lightweight, flexible Peripheral Event Linking System (PELS) with microcode programmable event-linking.
\Gls{pels} supports both single-wire event lines (instant actions) to optimize for stringent low-latency requirements and traditional commands issued through the system interconnect bus targeting peripheral configuration registers (sequenced actions). 
We integrate it in an \gls{ulp} RISC-V \gls{iot} processor and show that the power consumption of a linking event is reduced by 2.5 times compared to approaches that rely on the main core for the event-linking process at a low silicon area of \SI{7}{\kGE} in its minimal configuration when instantiated in a ULP RISC-V IoT processor.

\section*{Acknowledgments}
This work is supported by the dAIEDGE (101120726) and TRISTAN (101095947) projects, funded by HORIZON-CL4-2022-HUMAN-02 and HORIZON-CHIPS-JU, respectively.






\bibliographystyle{bibliography/IEEEtran}
\bibliography{bibliography/main}

\begin{thebibliography}{10}
\providecommand{\url}[1]{#1}
\csname url@samestyle\endcsname
\providecommand{\newblock}{\relax}
\providecommand{\bibinfo}[2]{#2}
\providecommand{\BIBentrySTDinterwordspacing}{\spaceskip=0pt\relax}
\providecommand{\BIBentryALTinterwordstretchfactor}{4}
\providecommand{\BIBentryALTinterwordspacing}{\spaceskip=\fontdimen2\font plus
\BIBentryALTinterwordstretchfactor\fontdimen3\font minus
  \fontdimen4\font\relax}
\providecommand{\BIBforeignlanguage}[2]{{%
\expandafter\ifx\csname l@#1\endcsname\relax
\typeout{** WARNING: IEEEtran.bst: No hyphenation pattern has been}%
\typeout{** loaded for the language `#1'. Using the pattern for}%
\typeout{** the default language instead.}%
\else
\language=\csname l@#1\endcsname
\fi
#2}}
\providecommand{\BIBdecl}{\relax}
\BIBdecl

\bibitem{EDGE_IOT_SURVEY_2023}
S.~Lu, J.~Lu, K.~An, X.~Wang, and Q.~He, ``Edge computing on iot for machine
  signal processing and fault diagnosis: A review,'' \emph{IEEE Internet of
  Things Journal}, pp. 1--1, 2023.

\bibitem{ENVISION}
B.~Moons, R.~Uytterhoeven, W.~Dehaene, and M.~Verhelst, ``14.5 envision: A
  0.26-to-10tops/w subword-parallel dynamic-voltage-accuracy-frequency-scalable
  convolutional neural network processor in 28nm fdsoi,'' in \emph{IEEE ISSCC},
  2017, pp. 246--247.

\bibitem{SENSOR_FUSION_SURVEY}
R.~M. Abdelmoneem, E.~Shaaban, and A.~Benslimane, ``{A Survey on Multi-Sensor
  Fusion Techniques in IoT for Healthcare},'' in \emph{2018 13th ICCES}, 2018,
  pp. 157--162.

\bibitem{SURVEILLANCE_IOT}
Y.~He, J.~Guo, and X.~Zheng, ``From surveillance to digital twin: Challenges
  and recent advances of signal processing for industrial internet of things,''
  \emph{IEEE S. P. Magazine}, vol.~35, no.~5, pp. 120--129, 2018.

\bibitem{RENESAS_AUTOMOTIVE_GATEWAY}
K.~Shimada, K.~Sano, K.~Fukuoka, H.~Morita, M.~Daito, T.~Kamei, H.~Hamasaki,
  and Y.~Shimazaki, ``A 33kdmips 6.4w vehicle communication gateway processor
  achieving 10gbps/w network routing, 40ms can bus start-up and 1.4mw standby
  power,'' in \emph{ISSCC}, 2023, pp. 240--242.

\bibitem{SENSOR_FUSION_OBJREC}
P.~Saha and S.~Mukhopadhyay, ``{Multispectral Information Fusion With
  Reinforcement Learning for Object Tracking in IoT Edge Devices},'' \emph{IEEE
  Sensors Journal}, vol.~20, no.~8, pp. 4333--4344, 2020.

\bibitem{SENSOR_FUSION_REDUNDANCY}
T.~He, L.~Zhang, F.~Kong, and A.~Salekin, ``Exploring inherent sensor
  redundancy for automotive anomaly detection,'' in \emph{DAC}, 2020, pp. 1--6.

\bibitem{SENSOR_FUSION_AUTOMOTIVE}
J.~Steinbaeck, C.~Steger, G.~Holweg, and N.~Druml, ``{Next generation radar
  sensors in automotive sensor fusion systems},'' in \emph{SDF}, 2017.

\bibitem{POWERNAP}
D.~Rossi, I.~Loi, A.~Pullini, C.~Müller, A.~Burg, F.~Conti, L.~Benini, and
  P.~Flatresse, ``A self-aware architecture for pvt compensation and power nap
  in near threshold processors,'' \emph{IEEE Design \& Test}, vol.~34, no.~6,
  pp. 46--53, 2017.

\bibitem{VEGA}
D.~Rossi, F.~Conti, M.~Eggiman, A.~D. Mauro, G.~Tagliavini, S.~Mach,
  M.~Guermandi, A.~Pullini, I.~Loi, J.~Chen, E.~Flamand, and L.~Benini, ``Vega:
  A ten-core soc for iot endnodes with dnn acceleration and cognitive wake-up
  from mram-based state-retentive sleep mode,'' \emph{IEEE JSSC}, vol.~57,
  no.~1, pp. 127--139, 2022.

\bibitem{UDMA}
A.~Pullini, D.~Rossi, G.~Haugou, and L.~Benini, ``{$\mu$DMA: An autonomous I/O
  subsystem for IoT end-nodes},'' in \emph{2017 27th PATMOS}, 2017, pp. 1--8.

\bibitem{PULPISSIMO}
P.~D. Schiavone, D.~Rossi, A.~Pullini, A.~Di~Mauro, F.~Conti, and L.~Benini,
  ``{Quentin: an Ultra-Low-Power PULPissimo SoC in 22nm FDX},'' in \emph{2018
  IEEE S3S}, 2018, pp. 1--3.

\bibitem{SILABS_PRS}
\BIBentryALTinterwordspacing
\emph{AN0025: Peripheral Reflex System}, Silicon Labs. [Online]. Available:
  \url{https://www.silabs.com/documents/public/application-notes/an0025-efm32-prs.pdf}
\BIBentrySTDinterwordspacing

\bibitem{RENESAS_LELC}
\BIBentryALTinterwordspacing
``New functions of the rl78/g23 logic and event link controller (lelc),'' Tech.
  Rep. [Online]. Available:
  \url{https://www.renesas.com/us/en/document/whp/new-functions-rl78g23-logic-and-event-link-controller-elcl?language=en}
\BIBentrySTDinterwordspacing

\bibitem{MICROCHIP_EVSYS}
\BIBentryALTinterwordspacing
\emph{Getting Started with Core Independent Peripherals on AVR}, Microchip.
  [Online]. Available:
  \url{{https://ww1.microchip.com/downloads/en/Appnotes/Microchip%20Getting%20Started%20with%20Core%20Independent%20Peripherals%20DS00002451B.pdf}}
\BIBentrySTDinterwordspacing

\bibitem{NORDIC_PPI}
\BIBentryALTinterwordspacing
\emph{PPI - Programmable Peripheral Interconnect}, Infineon. [Online].
  Available: \url{https://infocenter.nordicsemi.com/topic/ps_nrf52810/ppi.html}
\BIBentrySTDinterwordspacing

\bibitem{STM_PIM}
\BIBentryALTinterwordspacing
\emph{AN4640 Application note: Peripherals interconnections on STM32F4x
  series}, STMicroelectronics. [Online]. Available:
  \url{https://www.st.com/content/ccc/resource/technical/document/application_note/59/ed/30/07/55/76/4e/82/DM00154959.pdf/files/DM00154959.pdf/jcr:content/translations/en.DM00154959.pdf}
\BIBentrySTDinterwordspacing

\bibitem{NXP_XGATE}
\BIBentryALTinterwordspacing
``Tutorial: Introducing the xgate module to consumer and industrial application
  developers,'' Tech. Rep. [Online]. Available:
  \url{https://www.nxp.com.cn/docs/en/application-note/AN3224.pdf}
\BIBentrySTDinterwordspacing

\bibitem{BJORN_EVENT}
R.~A. Bjørnerud, M.~W. Lund, and K.~Svarstad, ``Event control and programming
  for microprocessor peripheral systems,'' in \emph{NORCHIP}, 2009, pp. 1--4.

\bibitem{SCM}
\BIBentryALTinterwordspacing
A.~Teman, D.~Rossi, P.~Meinerzhagen, L.~Benini, and A.~Burg, ``Power, area, and
  performance optimization of standard cell memory arrays through controlled
  placement,'' \emph{ACM Trans. Des. Autom. Electron. Syst.}, vol.~21, no.~4,
  may 2016. [Online]. Available: \url{https://doi.org/10.1145/2890498}
\BIBentrySTDinterwordspacing

\bibitem{PULP_NEAR_THRESHOLD}
M.~Gautschi, P.~D. Schiavone, A.~Traber, I.~Loi, A.~Pullini, D.~Rossi,
  E.~Flamand, F.~K. Gürkaynak, and L.~Benini, ``Near-threshold risc-v core
  with dsp extensions for scalable iot endpoint devices,'' \emph{IEEE TVLSI},
  vol.~25, no.~10, pp. 2700--2713, 2017.

\bibitem{IBEX}
\BIBentryALTinterwordspacing
lowRISC, \emph{lowRISC Ibex Reference Guide}, lowRISC. [Online]. Available:
  \url{https://ibex-core.readthedocs.io/en/latest/03_reference/index.html}
\BIBentrySTDinterwordspacing

\end{thebibliography}

\end{document}